\documentclass[a4paper]{jpconf}
\usepackage{graphicx}
\begin{document}
\title{Experimental investigation of tunneling times using Bose-Einstein condensates}

\author{Donatella Ciampini$^{1,2}$, Oliver Morsch$^2$ and Ennio Arimondo$^{1,2}$}

\address{$^1$ CNISM UdR Pisa, Dipartimento di Fisica 'E. Fermi', Lgo Pontecorvo 3, I-56127 Pisa, Italy}
\address{$^2$ INO-CNR, Dipartimento di Fisica 'E. Fermi', Lgo Pontecorvo 3, I-56127 Pisa, Italy}

\ead{ciampini@df.unipi.it}

\begin{abstract}
The time it takes a quantum system to complete a tunneling event (which in the case of cross-barrier tunneling can be viewed as the time spent in a classically forbidden area) is related to the time required for a state to evolve to an orthogonal state, and an observation, i.e., a quantum mechanical projection on a particular basis, is required to distinguish one state from another. We have performed time-resolved measurements of Landau-Zener tunneling of Bose-Einstein condensates in accelerated optical lattices, clearly resolving the steplike time dependence of the band populations. The use of different protocols enabled us to access the tunneling probability, in two different bases, namely, the adiabatic basis and the diabatic basis. The adiabatic basis corresponds to the eigenstates of the lattice, and the diabatic one to the free-particle momentum eigenstates. Our findings pave the way towards more quantitative studies of the tunneling time for LZ transitions, which are of current interest in the context of optimal quantum control and the quantum speed limit.
\end{abstract}

\section{Introduction}
Tunneling is one of the hallmarks of quantum systems, and physical effects associated with quantum tunneling are important in many branches of science~\cite{razavy_03}. While the probability for quantum tunneling can be readily calculated for a variety of systems and has been measured experimentally to great accuracy, the time it takes a quantum system to complete a tunneling event is a much less well-defined notion. It is often hard to measure in experiment and in many cases still the subject of intense debate~\cite{schulman_08}. In this paper we address the problem of measuring the timescale associated with one of the conceptually simplest tunneling phenomena, Landau-Zener (LZ) tunneling.

Landau-Zener tunneling arises when two energy levels of a quantum system cross as a function of some parameter that varies in time. There is a possibility of a transition if the degeneracy at the level crossing is lifted by a coupling and the system is forced across the resulting avoided crossing by varying the parameter that determines the level separation. LZ tunneling was first studied theoretically in the early 1930's in the context of atomic scattering processes and spin dynamics in time-dependent fields~\cite{Landau,Zener,Stueckelberg,Majorana32}

In its basic form the LZ problem can be described by a simple two-state model with a Hamiltonian given by
\begin{equation}
  H_{\rm LZ}=\left(
    \begin{array}{cc}
         \alpha t & \Delta E/2 \\
          \Delta E/2 &  -\alpha t\\
    \end{array}
  \right) \,
  \label{eqno1}
\end{equation}

 \begin{figure}[htc]
 \begin{center}
 \includegraphics[width=0.5\linewidth,angle=0]{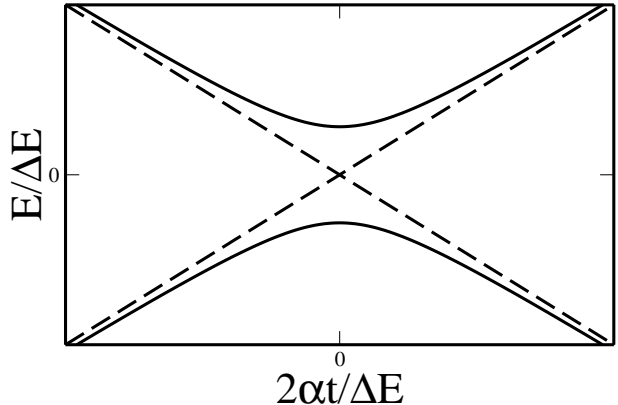}
 \caption{ \small \label{fig:1} Energy levels as a function of time. The dashed lines show the so-called diabatic levels i.e., the energy position of states in the absence of the interaction. The solid lines demonstrate the so-called adiabatic levels, i.e., the eigenstates of the system corresponding to the instantaneous Hamiltonian.}
 \end{center}
 \end{figure}
\noindent where the off-diagonal term, $\Delta E/2$, is the coupling between the two states and $\alpha$ is the rate of change of the energy levels in time. The dynamics of the system can be described either in the {\it diabatic} or in the {\it adiabatic} basis.  The diabatic basis is the basis of the bare states of Eq. (\ref{eqno1}) when there is coupling (i.e., no off-diagonal entries in the matrix). The adiabatic basis, on the other hand, is the basis of a system with a finite coupling $\Delta E/2$ between the two states. The Hamiltonian has two adiabatic energy levels $E_{\pm}=\pm\frac{1}{2}\sqrt{\left(2\alpha t\right)^2 + \Delta E ^{2}}$.\\

Assuming that the system is initially, at $t\rightarrow -\infty$, in the ground energy level $E_{\rm -}$ and if the sweeping rate is small enough, it will be exponentially likely that the system remains in its adiabatic ground state $E_{\rm -}$ at $t\rightarrow +\infty$. The limiting value of the adiabatic  LZ survival
probabilities (for $t$ going from $-\infty$ to $-\infty$)  is  \cite{Holthaus00},
\begin{equation}
{P_{\rm a}(\infty)=1-\exp\left(-\frac{\pi}{\gamma}\right),
\label{eqno2}}
\end{equation}
where we have introduced the dimensionless adiabaticity parameter $\gamma=4\hbar\alpha/\Delta E^{2}$.

While the above analysis can predict the probability for LZ tunneling very accurately, in contains no reference to the dynamics around the avoided crossing and hence to the time it takes the system to complete the tunneling event which eventually results in the populations measured in the ground and excited bands (or in the corresponding diabatic states). Also, in the case of LZ tunneling, which occurs in an abstract space spanned by the energy levels of the system as a function of a parameter, the concept of tunneling time is less intuitive than in the case of cross-barrier tunneling in real space, to name just one example. Nevertheless, the tunneling time (or transition time or jump time, as it is called in certain contexts) associated with LZ is a meaningful concept referring to the timescale on which the system evolves around the avoided crossing. Analytical estimates for the LZ transition times have been derived in~\cite{vitanov_96,vitanov_99} using the two-state model of Eq.~\ref{eqno1}. In a given basis, e.g., adiabatic or diabatic, different transition times  are obtained. Vitanov~\cite{vitanov_99} calculated the time-dependent diabatic/adiabatic survival probability at finite times. Analytical estimates for the LZ transition times were derived in~\cite{vitanov_96} using some exact and approximate results for the transition probability.

 In general, the LZ jump time in a given basis can be defined as the time after which the transition probability reaches its asymptotic value. From this definition one can expect to observe a step-like structure, with a finite width, in the time-resolved tunneling probability. Because the step is not very sharp,  it is not straightforward to define the initial and final times for the transition. It is even less obvious how to define the jump time for both small and for large coupling. Some possible choices have been used by Lim and Berry~\cite{Berry_90} and Vitanov~\cite{vitanov_96, vitanov_99}. The problem is even more complicated when the survival probability shows an oscillatory behavior on top of the step structure.  The oscillations give rise to an additional time scale in the system, namely an oscillation time and a damping time of the oscillations appearing in the transition probability after the crossing. Therefore, a measurement of the tunneling time depends very much on how these times are defined and also which basis is considered.

In~\cite{vitanov_99} the jump time in the diabatic/adiabatic bases is defined as
\begin{equation}
\tau^{jump}_{\rm d/a}=\frac{P_{\rm d/a}(\infty)}{P'_{\rm d/a}(0)}.
\label{eqno10}
\end{equation}
where $P_\mathrm{d/a}$ is the transition probability between the two diabatic/adiabatic states, respectively. $P'_{\rm d/a}(0)$  denotes the time derivative of the
tunneling probability  evaluated at the crossing
point. From this definition, the {\em diabatic} jump time is calculated as $\tau^{jump}_{\rm d} \approx \sqrt{2\pi\hbar/\alpha}$ is almost constant for large values of the  adiabaticity parameter $\gamma$. For $\gamma\ll1$, on the other hand, it decreases with $\gamma$ , $\tau^{jump}_{\rm d}  \approx 2\sqrt{\hbar(\gamma\alpha)^{-1}}$. In the {\em adiabatic} basis, when $\gamma$ is large, the transition probability is similar to that in the diabatic basis. For a small adiabaticity parameter, because of the oscillations appearing on top of the transition probability step structure, it is not straightforward to define the initial and the final time for the transition. Vitanov defines the initial jump time as the time $t<0$ at which the transition probability is very small (i.e., $P_{\rm a}(\tau)=\varepsilon P_{\rm a}(\infty)$, where $\varepsilon$ is a proper small number). The final time of the transition $t>0$ is defined as the time at which the non-oscillatory part of $P_{\rm a}(\tau)$ is equal to $(1+\varepsilon)P_{\rm a}(\infty)$. Using these definitions, Vitanov derived  that the transition time  in the adiabatic basis depends exponentially on the adiabaticity parameter, $\tau^{jump}_{\rm a} \approx \left(4/\epsilon\right)^{1/6}\gamma^{-1/3}\mathrm{exp}\left(\pi/(6\gamma)\right)\sqrt{\hbar/\alpha}$.

\section{Experimental results}
\label{results}
The Landau-Zener model is realized in our experiments using Bose-condensed rubidium atoms inside an optical lattice \cite{JonaLasinio03,Zenesini09}. Initially, we created Bose-Einstein condensates of
$5\times 10^4$ rubidium-87 atoms inside an optical dipole trap
(mean trap frequency around $80\,\mathrm{Hz}$). A one-dimensional
optical lattice created by two counter-propagating, linearly
polarized gaussian beams was then superposed on the BEC by ramping
up the power in the lattice beams in $100\,\mathrm{ms}$. The
wavelength of the lattice beams was $\lambda=842\,\mathrm{nm}$,
leading to a sinusoidal potential with lattice constant
$d_{\rm L}=\lambda/2=421\,\mathrm{nm}$.
A small frequency offset $\Delta \nu(t)$ between
the two beams could be introduced through the acousto-optic
modulators in the setup, which allowed us to accelerate the lattice in
a controlled fashion and hence, in the rest-frame of the lattice, to subject the atoms to a force $F_{\rm LZ}=Ma_\mathrm{LZ}$ with $a_\mathrm{LZ}=d_\mathrm{L}\frac{d\Delta \nu(t)}{dt}.$

The energy level structure of Bose condensates in optical lattices can be represented by energy bands in the Brillouin zone picture. At the edge of the Brillouin zone successive bands are separated by gaps, and in the vicinity of the zone edge our system approximates the LZ model very well. We can make time-resolved measurements of a single tunneling event in the following way: First, the Bose condensate is loaded adiabaticallly into a lattice, after which the lattice is accelerated to some finite quasimomentum. Thereafter, the instantaneous populations of the eigenstates of the system are measured, the exact protocol depending on the basis chosen. In detail, the protocols are as follows:
\begin{itemize}
\item

For measurements in the {\em adiabatic} basis, after loading the BEC into the optical lattice the lattice was accelerated with acceleration $a_\mathrm{LZ}$ for a time $t_\mathrm{LZ}$. The lattice thus acquired a final velocity
$v=a_\mathrm{LZ} t_\mathrm{LZ}$. At time $t=t_{\mathrm LZ}$ the acceleration was abruptly reduced to a smaller value
$a_\mathrm{sep}$ and the lattice depth was increased to
$V_\mathrm{sep}$ in a time $t_\mathrm{ramp}\ll T_\mathrm{B}$.
These values were chosen in such a way that at time
$t=t_\mathrm{LZ}$ the probability for LZ tunneling from
the lowest to the first excited energy band dropped from between
$\approx 0.1-0.9$ (depending on the initial parameters chosen) to
less than $\approx 0.01$, while the tunneling probability from the
first excited to the second excited band remained high at about
$0.95$. This meant that at $t=t_\mathrm{LZ}$ the tunneling process
was effectively interrupted and for $t>t_\mathrm{LZ}$ the measured
survival probability $P(t)=N_0/N_\mathrm{tot}$ (calculated from
the number of atoms $N_0$ in the lowest band and the total number
of atoms in the condensate $N_\mathrm{tot}$) reflected the
instantaneous value $P(t=t_\mathrm{LZ})$.

The lattice was then further accelerated for a time
$t_\mathrm{sep}$ such that $a_\mathrm{sep}t_\mathrm{sep}\approx 2n
p_\mathrm{rec}/M$ (where typically $n=2$ or $3$). In this way,
atoms in the lowest band were accelerated to a final velocity
$v\approx 2n p_\mathrm{rec}/M$, while atoms that had undergone
tunneling to the first excited band before $t=t_\mathrm{LZ}$
underwent further tunneling to higher bands with a probability
$>0.95$ and were, therefore, no longer accelerated. At time
$t_\mathrm{sep}$ the lattice and dipole trap beams were suddenly
switched off and the expanded atomic cloud was imaged after
$23\,\mathrm{ms}$. In these time-of-flight images the two velocity
classes $0$ and $2n p_\mathrm{rec}/M$ were well separated and the
atom numbers $N_0$ and $N_\mathrm{tot}$ could be measured
directly. Since the populations were effectively "frozen" inside the energy
bands of the lattice, which represent the adiabatic eigenstates of
the total Hamiltonian of the system, this experiment measured the time dependence of the LZ survival probability $P_{\rm a}$
in the {\it adiabatic} basis.

\item

For measurements in the {\em diabatic} basis, after the initial loading phase the lattice was accelerated with acceleration $a_\mathrm{LZ}$ for a time $t_\mathrm{LZ}$ as in the adiabatic case. At that point, however, the atomic sample was projected onto the free-particle diabatic basis by instantaneously (within less than $1\,\mathrm{\mu s}$) switching off the optical lattice. After
a time-of-flight the number of atoms in the $v=0$ and
$v=2p_\mathrm{rec}/M$ momentum classes are measured and from these
the survival probability (corresponding to the atoms remaining in
the $v=0$ velocity class relative to the total atom number) is
calculated.
\end{itemize}

\begin{figure}[htc]
 \begin{center}
 \includegraphics[width=0.7\linewidth,angle=0]{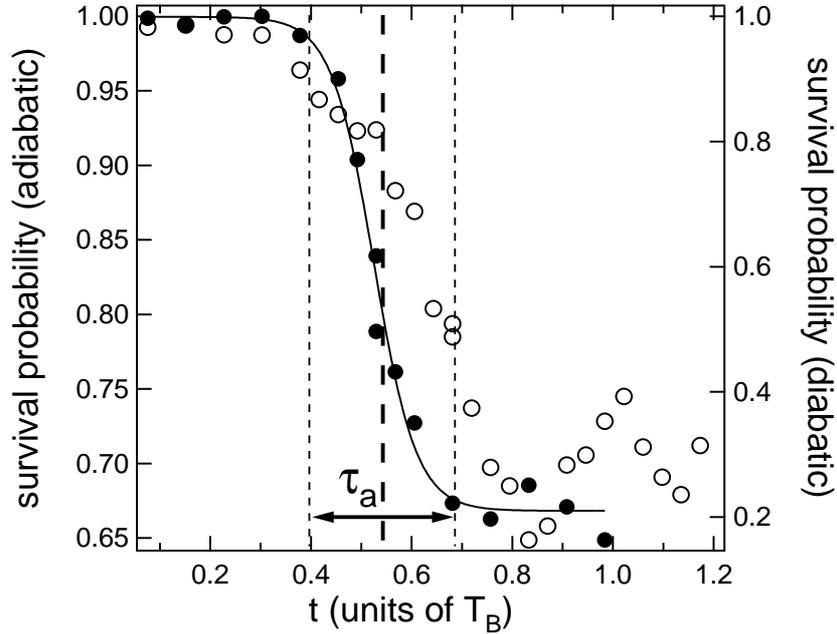}
 \caption{ \small \label{fig:experiment} Time-resolved measurements of LZ tunneling in the adiabatic (filled circles) and diabatic bases (open circles)for $F_0=1.197$ and $V/E_\mathrm{rec}=1.8$. The solid line is a sigmoid fit to the adiabatic survival probability, and the vertical dashed lines indicate the position of the zone edge at $t=0.5\,T_B$ and the tunneling time $\tau_a$.}
 \end{center}
 \end{figure}

The results of typical measurements in the adiabatic and diabatic bases are shown in Fig. \ref{fig:experiment}. The step-like behaviour of the survival probability around $t=0.5T_B$ is clearly visible, which demonstrates that our experimental protocol does, indeed, allow us to access the timescale of the LZ transition. It is also obvious from the figure that while in the adiabatic basis the transition is smooth and can be well fitted with a sigmoid function, in the diabatic basis there are strong oscillations for times $t>0.5\,T_B$. As a consequence, the tunneling time in the adiabatic basis can be easily identified as the width of the transition curve (indicated in the figure), while in the diabatic basis it is less obvious when the tunneling event is completed.

\section{Conclusions}
We have demonstrated that experiments with Bose-condensates in accelerated optical lattices allow access to the full dynamics of LZ tunneling and hence to the timescales involved in the tunneling process, both in the adiabatic and diabatic bases of the problem. Our experiments pave the way towards a thorough investigation of tunneling times and the quantum speed limit. The latter has recently been discussed theoretically in the context of optimal quantum control \cite{QSL}QSL and more generalized Landau-Zener protocols involving non-linear sweep functions that are predicted to lead to shorter minimum times for completing a tunneling event. As our experimental setup allows us to realize arbitrary protocols for the lattice acceleration, such experiments should be relatively straightforward to realize.

\ack
The authors would like to thank the PhD students and post-docs participating in the experiments described in this article: A. Zenesini, H. Lignier and J. Radogostowicz. We acknowledge funding by the EU Project "NAMEQUAM" (EC FP7-225187) and the CNISM "Progetto Innesco 2007".

\end{document}